\documentclass[traditabstract, twocolumns, longauth]{aa}

\usepackage{graphicx}
\usepackage{txfonts}
\usepackage{natbib}
\usepackage{xspace}
\usepackage{color}
\usepackage{hyperref}
\bibpunct{(}{)}{;}{a}{}{,}

\newcommand{\obj}{DES~J0408$-$5354}

\usepackage{eso-pic}
\AddToShipoutPictureBG*{%
\AtPageUpperLeft{%
\hspace{0.7\paperwidth}%
\raisebox{-4.3\baselineskip}{%
\makebox[0pt][l]{\textnormal{DES-2017-0252}}
}}}%
\AddToShipoutPictureBG*{%
\AtPageUpperLeft{%
\hspace{0.7\paperwidth}%
\raisebox{-5.5\baselineskip}{%
\makebox[0pt][l]{\textnormal{Fermilab PUB-17-204-AE}}
}}}%

\begin{document}

\title{COSMOGRAIL XVI: Time delays for the quadruply imaged quasar \obj\ with high-cadence photometric monitoring}
\titlerunning{Time delay measurement in \obj}

\author{
F. Courbin \inst{\ref{epfl}} \and
V. Bonvin \inst{\ref{epfl}} \and 
E. Buckley-Geer \inst{\ref{fermilab}} \and
C.D. Fassnacht \inst{\ref{ucdavis}} \and
J. Frieman \inst{\ref{fermilab}, \ref{chicago}} \and
H. Lin \inst{\ref{fermilab}} \and
P.J. Marshall \inst{\ref{stanford}} \and
S.H. Suyu \inst{\ref{MPG}, \ref{TUM}, \ref{ASIAA}} \and
T. Treu \inst{\ref{ucla}} \and
T. Anguita \inst{\ref{UNAB}, \ref{millenium}} \and
V. Motta \inst{\ref{valpo}} \and
G. Meylan \inst{\ref{epfl}} \and
E. Paic \inst{\ref{epfl}} \and
M. Tewes \inst{\ref{bonn}} \and
A. Agnello \inst{\ref{eso}} \and
D.C.-Y. Chao \inst{\ref{MPG}} \and
M. Chijani \inst{\ref{UNAB}} \and
D. Gilman \inst{\ref{ucla}} \and 
K. Rojas \inst{\ref{valpo}} \and
P. Williams \inst{\ref{ucla}} \and
A. Hempel \inst{\ref{UNAB}} \and 
S. Kim \inst{\ref{puc}, \ref{heidelberg}}\and
R. Lachaume \inst{\ref{puc}, \ref{heidelberg}}\and 
M. Rabus \inst{\ref{puc}, \ref{heidelberg}}\and
T. M. C.~Abbott\inst{\ref{ctio}} \and
S.~Allam\inst{\ref{fermilab}} \and
J.~Annis\inst{\ref{fermilab}} \and
M.~Banerji\inst{\ref{camastro}, \ref{camkavli}} \and
K.~Bechtol\inst{\ref{lsst}} \and
A.~Benoit-L{\'e}vy\inst{\ref{cnrs}, \ref{ucl}, \ref{sorb}} \and
D.~Brooks\inst{\ref{ucl}} \and
D.~L.~Burke\inst{\ref{stanford}, \ref{slac}} \and
A. Carnero Rosell\inst{\ref{linea}, \ref{obsnat}} \and
M.~Carrasco~Kind\inst{\ref{uiuc}, \ref{ncsa}} \and
J.~Carretero\inst{\ref{ifae}} \and
C.~B.~D'Andrea\inst{\ref{upenn}} \and
L.~N.~da Costa\inst{\ref{linea}, \ref{obsnat}} \and
C.~Davis\inst{\ref{stanford}} \and
D.~L.~DePoy\inst{\ref{tam}} \and
S.~Desai\inst{\ref{kandi}} \and
B.~Flaugher\inst{\ref{fermilab}} \and
P.~Fosalba\inst{\ref{ieec}} \and
J.~Garc\'ia-Bellido\inst{\ref{uam}} \and
E.~Gaztanaga\inst{\ref{ieec}} \and
D.~A.~Goldstein\inst{\ref{ucb}, \ref{lbnl}} \and
D.~Gruen\inst{\ref{stanford}, \ref{slac}} \and
R.~A.~Gruendl\inst{\ref{uiuc}, \ref{ncsa}} \and
J.~Gschwend\inst{\ref{linea}, \ref{obsnat}} \and
G.~Gutierrez\inst{\ref{fermilab}} \and
K.~Honscheid\inst{\ref{caap}, \ref{physosu}} \and
D.~J.~James\inst{\ref{uwash}, \ref{ctio}} \and
K.~Kuehn\inst{\ref{aao}} \and
S.~Kuhlmann\inst{\ref{anl}} \and
N.~Kuropatkin\inst{\ref{fermilab}} \and
O.~Lahav\inst{\ref{ucl}} \and
M.~Lima\inst{\ref{saopaulo}, \ref{linea}} \and
M.~A.~G.~Maia\inst{\ref{linea}, \ref{obsnat}} \and
M.~March\inst{\ref{upenn}} \and
J.~L.~Marshall\inst{\ref{tam}} \and
R.~G.~McMahon\inst{\ref{camastro}, \ref{camkavli}} \and
F.~Menanteau\inst{\ref{uiuc}, \ref{ncsa}} \and
R.~Miquel\inst{\ref{icrea}, \ref{ifae}} \and
B.~Nord\inst{\ref{fermilab}} \and
A.~A.~Plazas\inst{\ref{caltecg}} \and
E.~Sanchez\inst{\ref{ciemat}} \and
V.~Scarpine\inst{\ref{fermilab}} \and
R.~Schindler\inst{\ref{slac}} \and
M.~Schubnell\inst{\ref{umich}} \and
I.~Sevilla-Noarbe\inst{\ref{ciemat}} \and
M.~Smith\inst{\ref{usouth}} \and
M.~Soares-Santos\inst{\ref{fermilab}} \and
F.~Sobreira\inst{\ref{uec}, \ref{linea}} \and
E.~Suchyta\inst{\ref{oakridge}} \and
G.~Tarle\inst{\ref{umich}} \and
D.~L.~Tucker\inst{\ref{fermilab}} \and
A.~R.~Walker\inst{\ref{ctio}} \and
W.~Wester\inst{\ref{fermilab}} }

\institute{
Institute of Physics, Laboratory of Astrophysics, Ecole Polytechnique 
F\'ed\'erale de Lausanne (EPFL), Observatoire de Sauverny, 1290 Versoix, 
Switzerland \label{epfl}\goodbreak \and
Fermi National Accelerator Laboratory, P.O. Box 500, Batavia, IL 60510, 
USA \label{fermilab}\goodbreak \and
Department of Physics, University of California, Davis, CA 95616, USA 
\label{ucdavis}\goodbreak \and
Kavli Institute for Cosmological Physics, University of Chicago, 
Chicago, IL 60637, USA \label{chicago}\goodbreak \and
Kavli Institute for Particle Astrophysics and Cosmology, Stanford 
University, 452 Lomita Mall, Stanford, CA 94035, USA 
\label{stanford}\goodbreak \and
Max Planck Institute for Astrophysics, Karl-Schwarzschild-Strasse
1, D-85740 Garching, Germany \label{MPG}\goodbreak \and
Physik-Department, Technische Universit\"at M\"unchen, 
James-Franck-Stra\ss{}e~1, 85748 Garching, Germany \label{TUM}\goodbreak 
\and
Institute of Astronomy and Astrophysics, Academia Sinica, P.O.~Box 
23-141, Taipei 10617, Taiwan \label{ASIAA}\goodbreak \and  
Department of Physics and Astronomy, University of California, Los 
Angeles, CA 90095, USA \label{ucla}\goodbreak \and
Departamento de Ciencias F\'isicas, Universidad Andres Bello Fernandez 
Concha 700, Las Condes, Santiago, Chile    \label{UNAB}\goodbreak \and
Millennium Institute of Astrophysics, Chile \label{millenium}\goodbreak 
\and
Instituto de F\'isica y Astronom\'ia, Universidad de Valpara\'iso, Avda. 
Gran Breta\~na 1111, Playa Ancha, Valpara\'iso 2360102, Chile 
\label{valpo}\goodbreak \and
Argelander-Institut f\"ur Astronomie, Auf dem H\"ugel 71, 53121, Bonn, 
Germany \label{bonn}\goodbreak
\and
European Southern Observatory, Karl-Schwarzschild-Strasse 2, D-85748 
Garching bei Munchen, Germany \label{eso}\goodbreak \and
Centro de Astroingenier\'ia, Facultad de F\'isica, Pontificia Universidad 
Cat\'olica de Chile, Av. Vicu\~na Mackenna 4860, Macul 7820436, 
Santiago, Chile \label{puc}\goodbreak \and
Max-Planck-Institut f\"ur Astronomie, K\"onigstuhl 17, 69117 Heidelberg, Germany \label{heidelberg}\goodbreak \and
Cerro Tololo Inter-American Observatory, National Optical Astronomy 
Observatory, Casilla 603, La Serena, Chile {\label{ctio}}\goodbreak \and
Institute of Astronomy, University of Cambridge, Madingley Road, 
Cambridge CB3 0HA, UK {\label{camastro}}\goodbreak \and
Kavli Institute for Cosmology, University of Cambridge, Madingley Road, 
Cambridge CB3 0HA, UK {\label{camkavli}}\goodbreak \and
LSST, 933 North Cherry Avenue, Tucson, AZ 85721, USA 
{\label{lsst}}\goodbreak \and
CNRS, UMR 7095, Institut d'Astrophysique de Paris, F-75014, Paris, 
France {\label{cnrs}}\goodbreak \and
Department of Physics \& Astronomy, University College London, Gower 
Street, London, WC1E 6BT, UK {\label{ucl}}\goodbreak \and
Sorbonne Universit\'es, UPMC Univ Paris 06, UMR 7095, Institut 
d'Astrophysique de Paris, F-75014, Paris, France 
{\label{sorb}}\goodbreak \and
Kavli Institute for Particle Astrophysics \& Cosmology, P. O. Box 2450, 
Stanford University, Stanford, CA 94305, USA 
{\label{stanford}}\goodbreak \and
SLAC National Accelerator Laboratory, Menlo Park, CA 94025, USA 
{\label{slac}}\goodbreak \and
Laborat\'orio Interinstitucional de e-Astronomia - LIneA, Rua Gal. 
Jos\'e Cristino 77, Rio de Janeiro, RJ - 20921-400, Brazil 
{\label{linea}}\goodbreak \and
Observat\'orio Nacional, Rua Gal. Jos\'e Cristino 77, Rio de Janeiro, RJ 
- 20921-400, Brazil {\label{obsnat}}\goodbreak \and
Department of Astronomy, University of Illinois, 1002 W. Green Street, 
Urbana, IL 61801, USA {\label{uiuc}}\goodbreak \and
National Center for Supercomputing Applications, 1205 West Clark St., 
Urbana, IL 61801, USA {\label{ncsa}}\goodbreak \and
Institut de F\'{\i}sica d'Altes Energies (IFAE), The Barcelona Institute 
of Science and Technology, Campus UAB, 08193 Bellaterra (Barcelona) 
Spain {\label{ifae}}\goodbreak \and
Department of Physics and Astronomy, University of Pennsylvania, 
Philadelphia, PA 19104, USA {\label{upenn}}\goodbreak \and
George P. and Cynthia Woods Mitchell Institute for Fundamental Physics 
and Astronomy, and Department of Physics and Astronomy, Texas A\&M 
University, College Station, TX 77843,  USA {\label{tam}}\goodbreak \and
Department of Physics, IIT Hyderabad, Kandi, Telangana 502285, India 
{\label{kandi}}\goodbreak \and
Institut de Ci\`encies de l'Espai, IEEC-CSIC, Campus UAB, Carrer de Can 
Magrans, s/n,  08193 Bellaterra, Barcelona, Spain 
{\label{ieec}}\goodbreak \and
Instituto de Fisica Teorica UAM/CSIC, Universidad Autonoma de Madrid, 
28049 Madrid, Spain {\label{uam}}\goodbreak \and
Department of Astronomy, University of California, Berkeley,  501 
Campbell Hall, Berkeley, CA 94720, USA {\label{ucb}}\goodbreak \and
Lawrence Berkeley National Laboratory, 1 Cyclotron Road, Berkeley, CA 
94720, USA {\label{lbnl}}\goodbreak \and
Center for Cosmology and Astro-Particle Physics, The Ohio State 
University, Columbus, OH 43210, USA {\label{caap}}\goodbreak \and
Department of Physics, The Ohio State University, Columbus, OH 43210, 
USA {\label{physosu}}\goodbreak \and
Astronomy Department, University of Washington, Box 351580, Seattle, WA 98195, USA {\label{uwash}} \and
Australian Astronomical Observatory, North Ryde, NSW 2113, Australia 
{\label{aao}}\goodbreak \and
Argonne National Laboratory, 9700 South Cass Avenue, Lemont, IL 60439, 
USA {\label{anl}}\goodbreak \and
Departamento de F\'isica Matem\'atica, Instituto de F\'isica, 
Universidade de S\~ao Paulo, CP 66318, S\~ao Paulo, SP, 05314-970, 
Brazil {\label{saopaulo}}\goodbreak \and
Instituci\'o Catalana de Recerca i Estudis Avan\c{c}ats, E-08010 Barcelona, Spain {\label{icrea}} \and
Jet Propulsion Laboratory, California Institute of Technology, 4800 Oak 
Grove Dr., Pasadena, CA 91109, USA {\label{caltecg}}\goodbreak \and
Centro de Investigaciones Energ\'eticas, Medioambientales y 
Tecnol\'ogicas (CIEMAT), Madrid, Spain {\label{ciemat}}\goodbreak \and
Department of Physics, University of Michigan, Ann Arbor, MI 48109, USA 
{\label{umich}}\goodbreak \and
School of Physics and Astronomy, University of Southampton,  
Southampton, SO17 1BJ, UK {\label{usouth}}\goodbreak \and
Instituto de F\'isica Gleb Wataghin, Universidade Estadual de Campinas, 
13083-859, Campinas, SP, Brazil {\label{uec}}\goodbreak \and
Computer Science and Mathematics Division, Oak Ridge National 
Laboratory, Oak Ridge, TN 37831 {\label{oakridge}\goodbreak}
}

\date{\today}
\abstract{We present time-delay measurements for the new quadruple 
imaged quasar \obj, the first quadruple imaged quasar found in the Dark 
Energy Survey (DES). Our result is made possible by implementing a new 
observational strategy using almost daily observations with the MPIA 
2.2m telescope at La Silla observatory and deep exposures reaching a 
signal-to-noise ratio of about 1000 per quasar image. This data quality 
allows us to catch small photometric variations (a few mmag 
rms) of the quasar, acting on temporal scales much shorter than 
microlensing, hence making the time delay measurement very robust 
against microlensing. In only 7 months we measure very accurately one of 
the time delays in \obj: $\Delta t(AB) = -112.1 \pm 2.1$ days (1.8\%) 
using only the MPIA 2.2m data. In combination with data taken with the 
1.2m Euler Swiss telescope, we also measure two delays involving the D 
component of the system $\Delta t(AD) = -155.5 \pm 12.8$ days (8.2\%) 
and $\Delta t(BD) = -42.4 \pm 17.6$ days (41\%), where all the error 
bars include systematics. Turning these time delays into cosmological 
constraints will require deep HST imaging or ground-based Adaptive 
Optics (AO), and information on the velocity field of the lensing 
galaxy.}


\maketitle

\begin{figure*}[t!]
\centering
\includegraphics[width=18.cm]{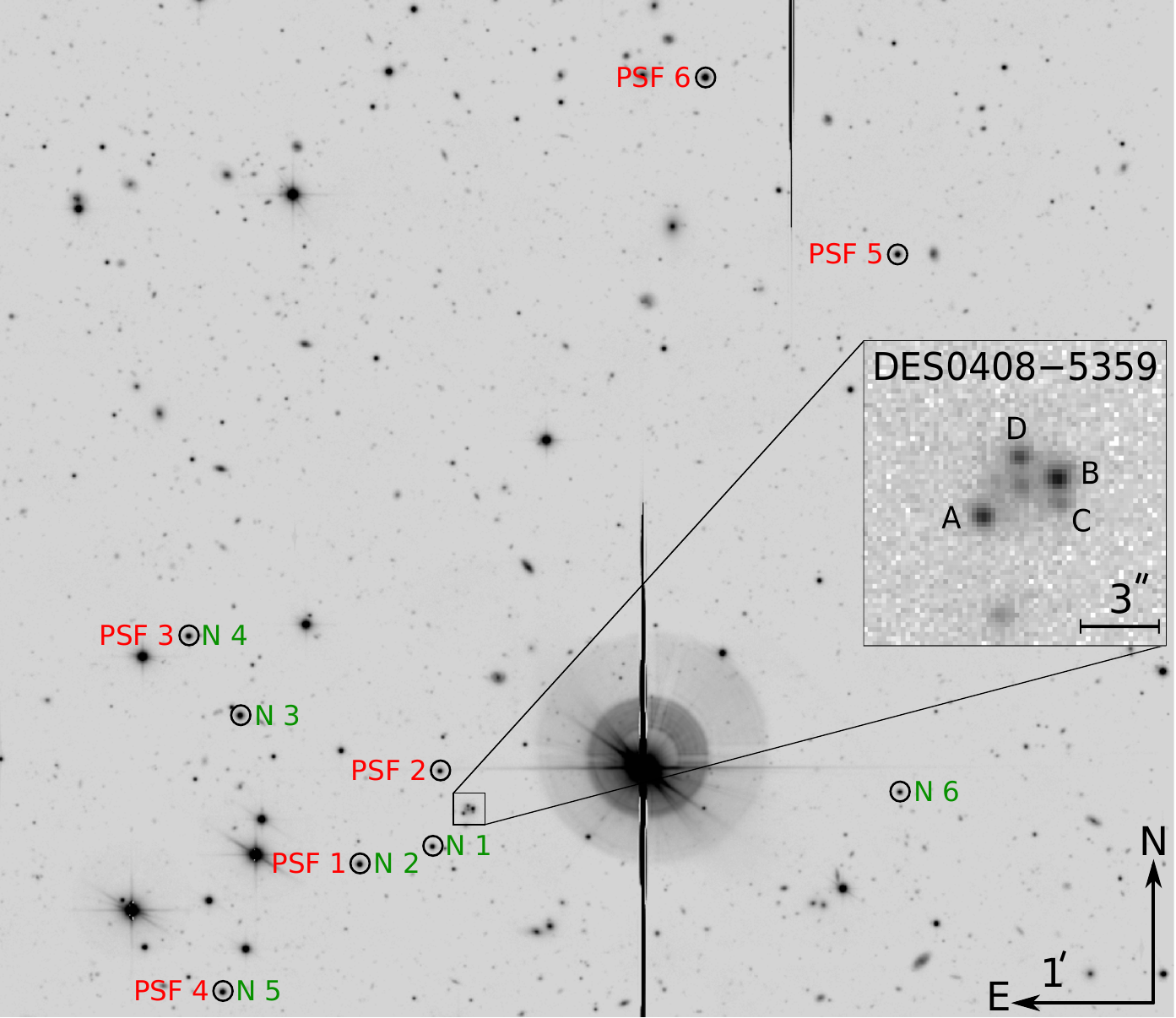}
\caption{Part of the field of view around \obj\, as seen with the 2.2m 
MPIA/ESO telescope at La Silla observatory. The image is a stack of 150 
frames totaling 25h of exposure. The 6 PSF stars used to obtain the 
photometric light curves with deconvolution photometry are indicated in 
red. The 6 stars used for the frame-to-frame calibration of the relative 
photometry are indicated in green. The inset shows a 10\arcsec\ zoom on 
\obj\ and is extracted from a single 640 sec exposure with 0.6\arcsec\ 
seeing with the same labeling for the quasar images as in Fig.~1 of 
\citet[][]{Lin2017}. Note that image C is blended with a foreground 
lensing galaxy labeled G2 in \citet[][]{Agnello2017} and 
\citet[][]{Lin2017}.}
\label{fig:field}
\end{figure*}
%

%
\section{Introduction}

Accurate and precise measurements of the time delay(s) between the multiple images of gravitationally lensed quasars offer an independent way of constraining cosmology. The method is simple and is mostly sensitive to $H_0$ with weak dependence on other cosmological parameters \citep{Refsdal1964}. For this reason, the time-delay method has the potential to alleviate the degeneracies between cosmological parameters other than $H_0$. In addition, it provides helpful input to resolve the tension between $H_0$ as measured by Planck assuming a flat $\Lambda$CDM model \citep{Planck2016} and the local distance ladder, i.e. Cepheid stars \citep{Freedman2017, Freedman2001} and type Ia supernovae \cite[e.g.][]{Riess2016}. Quasar time delays offer an opportunity to measure $H_0$ completely independently of any of the above probes.

The method requires several ingredients: i) time-delay measurements, ii) models constraining the mass and light distribution in the lensing galaxy, iii) an estimate of the contribution of objects along the line of sight to the overall potential well. The first point has been addressed by the COSMOGRAIL program, started in 2004 and delivering since then some of the best-quality time-delay measurements \citep[e.g.][]{Bonvin2017, Rathna2013, Tewes2013, Courbin2011, Vuissoz2008, Courbin2005, Eigenbrod2005}. In parallel, detailed modeling techniques have been developed and used on deep Hubble Space Telescope (HST) images in combination with spectroscopic data  providing crucial constraints on the dynamics of the lensing galaxy \citep{tt2002, Suyu2006, Suyu2009}. Such models, in combination with an estimate of the overall mass along the line-of-sight \cite[e.g.][]{Hilbert2009, McCully2017, McCully2014, Collett2013} allow one to measure the time-delay distance and consequently the Hubble parameter, $H_0$ \citep[e.g.][]{Suyu2010}. 

In order to perform precise cosmological measurement with strongly lensed quasars, these three ingredients must be accurately constrained. This has become possible only recently, with the joint efforts of the COSMOGRAIL \citep[e.g.][]{Courbin2005} and H0LiCOW programs \citep[$H_0$ Lenses in COSMOGRAIL's Wellspring;][]{Suyu2017}, focusing on 5 well selected bright lensed quasars. Recent results can be found in \citet{Bonvin2017, Wong2017, Rusu2017, Sluse2016} who infer $H_0 =71.9^{+2.4}_{-3.0}\ {\rm km\ s^{-1} Mpc^{-1}}$ from 3 of the H0LiCOW lenses in a flat $\Lambda$CDM universe.  
 
The H0LiCOW sample currently under study includes 5 lenses, with an 
expected $H_0$ measurement to $<$3.5\% including systematics 
\citep{Suyu2017}. Going beyond this will require mass production of time 
delays. With 55 new time delays and dynamical measurements for the 
lensing galaxy  \citet{Jee2015} estimate that $H_0$ can be measured to 
close to 1\%. An independent study by Shajib, Agnello \& Treu (2017, in 
prep.) show that with resolved kinematics of the lens (e.g. with JWST or 
ground-based AO) 1\% accuracy on $H_0$ can be reached with 40 lenses. 
This requires the discovery of new lenses, which is underway in the Dark 
Energy Survey (DES) 
\citep{Ostrovski2017,Lin2017,Agnello2015}, deep spectroscopy, the characterisation of the line-of-sight matter distribution, and the measurement of the time delays to a few percents for each individual system. The latter is the goal of the present work.

Because the slow intrinsic variations of the quasar occur roughly at the same time scale as the extrinsic variations (i.e. microlensing), measuring time delays requires years of monitoring. As the future of time delay cosmography resides in the measurement of several tens of new time delays, each time delay must be measured shortly after the start of the monitoring campaign, i.e. much faster than the typical 10 years it takes with current lens monitoring data. Current lens monitoring campaigns, including COSMOGRAIL, use 1m-class telescopes with a monitoring cadence of about 1 epoch per 3-4 days. The typical photometric accuracy with such data is limited to about 0.01 mag rms for many targets, hence allowing to catch only the most prominent features of the quasar variations. It is difficult, and sometimes impossible, to sufficiently  disentangle these from extrinsic variations related to microlensing unless very long light curves are available \cite[e.g.][]{BonvinPyCS2016, Liao2015}.

In the present work, we implement a new high-cadence and 
high-signal-to-noise (high-S/N) lens monitoring program, with the goal 
of measuring time delays in only 1 single observing season. With on 
average one observing point per day and a S/N of the order of 1000 per 
quasar component, we can now catch much faster variability in the 
intrinsic light curve of the quasar \citep[e.g.][]{Mosquera2011}. In 
almost all cases, these features on time scales of a few days to a few 
weeks are more than an order of magnitude faster than the extrinsic 
variations. This difference in signal frequencies makes it possible to 
disentangle much better between extrinsic and intrinsic quasar 
variations. As the small and fast quasar variations are frequent 
\citep[see, e.g. the Kepler data for AGNs in][]{Mushotzky2011}, only a 
short monitoring period is required to measure time delays, i.e. 
catching significant quasar variations is guaranteed in a 1-year period 
provided high S/N and high-cadence data are available. This achieved by 
using the MPIA 2.2m telescope and the Wide Field Imager (WFI) at ESO La 
Silla Observatory daily, through a dedicated monitoring program.

We present here our first time delay measurement obtained with the MPIA 
2.2m telescope, for the quadruply-imaged quasar \obj, at $z_q=2.375$. 
\obj\ was identified as a quadruple imaged quasar by \citet{Lin2017}. 
The lensing galaxy has a redshift of $z_l=0.597$, measured by 
\cite[][]{Lin2017} using the Gemini-South telescope. 
\citet[][]{Agnello2017} provide simple models for \obj\ using a deep 
image of the lens obtained from WFI data and predict time delays for a 
$\Lambda$CDM cosmology and different mass distributions including 
potential companions to the lensing galaxy, which influence the 
time-delay predictions. 

\begin{figure}[t!]
\centering
\includegraphics[width=8.7cm]{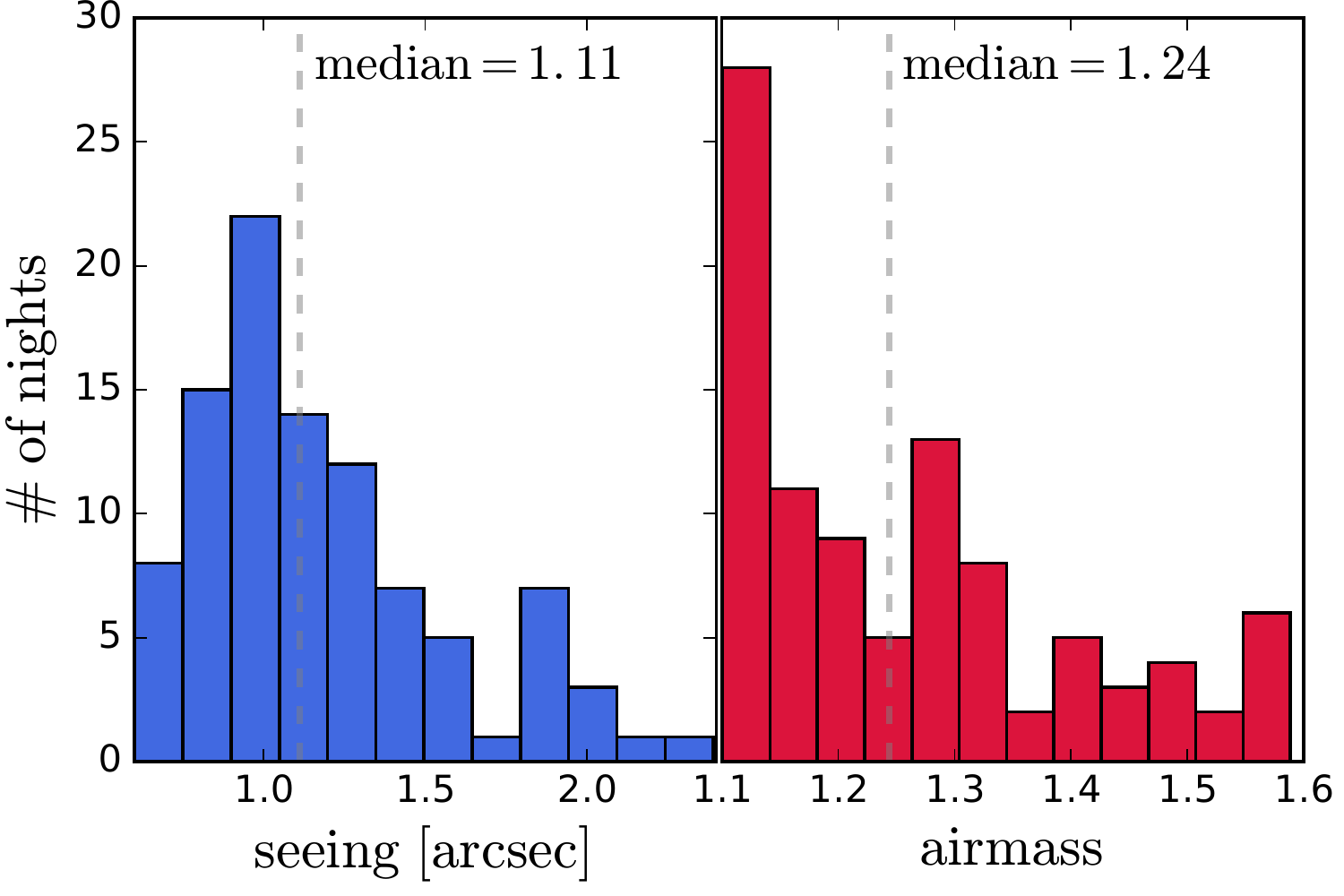}
\caption{Seeing and airmass distributions for the 7 months of observations of \obj\ with the WFI instrument on the MPIA 2.2m telescope.}
\label{fig:seeing}
\end{figure}

\begin{figure*}[t!]
\centering
\includegraphics[width=.98\textwidth]{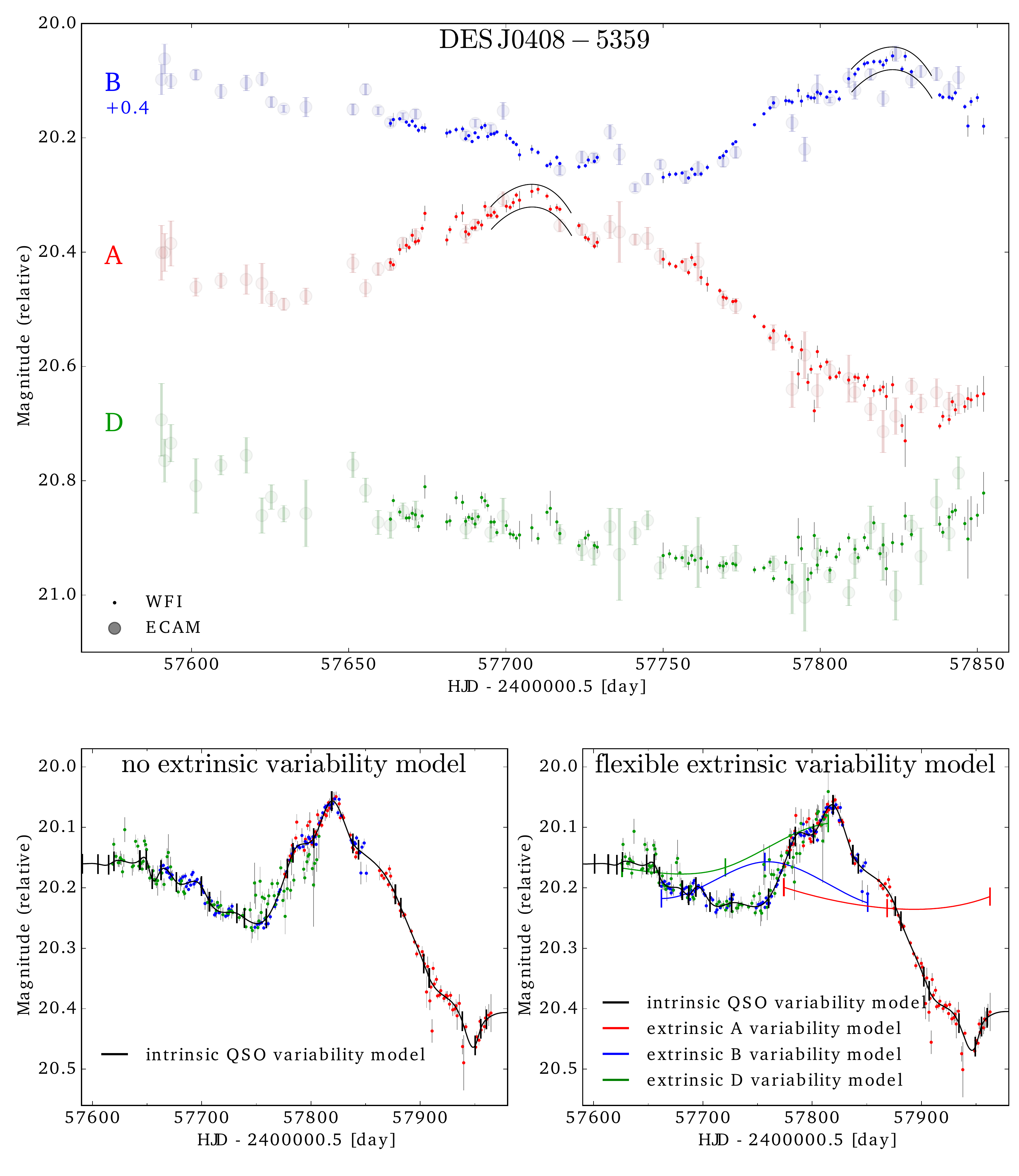}
\caption{{\it Top:} light curves for \obj\, obtained in the $R_c$ filter 
with the MPIA 2.2m telescope and the WFI instrument. The points obtained 
with the 1.2m Euler telescope are also shown, with larger and thicker 
symbols. To guide the eye, the structure constraining the most the AB 
time delay is indicated between black solid lines. {\it Bottom left:} 
spline-fitting of the intrinsic quasar variations (with an initial knot 
step of $\eta=15$ days) and time delay determination when neglecting 
extrinsic variation due to microlensing. Note that the time delay values 
do not depend much on the choice of this $\eta$ parameter which is only 
an initial value optimized during the fit. {\it Bottom right:} same as 
bottom left but now including extrinsic variations (color curves). For 
more clarity the lower S/N Euler data are shown only in the upper left 
panel. Our light curves are available online at CDS and on 
\url{cosmograil.org}.}
\label{fig:lc}
\end{figure*}

\section{Observations and photometry}

The observational material for the present time-delay measurement consists of almost daily imaging data with the MPIA 2.2m telescope and of bi-weekly imaging with the 1.2m Euler Swiss telescope, both at ESO La Silla. 

We started the observations on 1 October 2016 with the MPIA 2.2m telescope at ESO La Silla to monitor \obj\ through the $R_c$ filter. The WFI instrument, mounted in the 2.2m telescope, has a total field of view of 36\arcmin\ $\times$ 36\arcmin, covered by 8 CCDs with a pixel size of 0.2\arcsec. For our monitoring purpose we use only 1 chip to ensure a stable night-to-night calibration. This chip has a field of view of 9\arcmin\ $\times$ 18\arcmin. Part of this it is shown in Fig.~\ref{fig:field}.

The WFI was used almost daily until 8 April 2017, i.e. over a total of 7 
months of visibility of the object, except for 14 consecutive nights 
between 10 December 2016 and 24 December 2016 due to technical problems 
and for 1 week in January 2017 due to an extended period of poor 
weather. For each observing epoch, 4 dithered exposures of 640 sec each 
were taken in the $R_c$ filter. A total of 459 images were taken in 7 
months, of which we use 398 which have adequate seeing and PSF quality. 
More precisely, we removed the images with 1- a seeing above 3.0\arcsec, 
2- a mean ellipticity above $e=0.4$, 3- a sky level about 10000 
electrons, 4- obvious failure in the PSF modeling. On average, the 
resulting temporal sampling was one observing point every 1.96 day. The 
median seeing over this period was 1.1\arcsec. Thanks to the flexible 
scheduling of the observations at the telescope it was possible to 
observe \obj\ most of the time at low airmass. The seeing and airmass 
distributions of the observations are given in Fig.~\ref{fig:seeing}.

The high-cadence and high-S/N (2-3 mmag rms per quasar image) obtained with the 2.2m telescope allow us to catch much smaller and much shorter photometric variations than the COSMOGRAIL observations obtained with smaller 1m-size telescopes. We can typically see signals as small as a few mmags and as short as a 15-20 days, which is crucial to avoid contamination by extrinsic variations, as  illustrated in Sect.~\ref{sec:PyCS}

The data from the 1.2m Euler telescope were obtained in the $R$-band 
with the ECAM instrument from July 2016 to April 2017. The pixel size of 
the camera is 0.238\arcsec, providing a field of view of 14\arcmin\ 
on-a-side. We took, for each of the 45 observing epochs 6 exposures of 
360 sec each, i.e. 36 min in total. The mean temporal sampling for the 
Euler observations is of only 1 point every 5 days, but the Euler 
observations started about 100 days before the WFI observations, hence 
extending the length of the light curves. 

The data reduction procedure applied to the images follows the standard COSMOGRAIL pipeline, as applied to the data obtained with the 1.2m Euler telescope for RX~J1131$-$123 \citep{Tewes2013} and HE~0435$-$1223 \citep{Bonvin2017}. It includes subtraction of a bias level and flat-fielding using sky flats taken on average every few nights. Each frame is then sky-subtracted using the {\tt GLOBAL} option in the {\tt Sextractor} package \citep{Bertin1996}. The data from the 2.2m telescope have significant fringe patterns on bright nights. We therefore construct a fringe image by iteratively sigma-clipping the four dithered exposure of each night and by taking the median. This image is then subtracted from the individual dithered exposures taken each night which are subsequently registered to the same pixel grid. 

\begin{figure*}[t!]
\centering
\includegraphics[width=.98\textwidth]{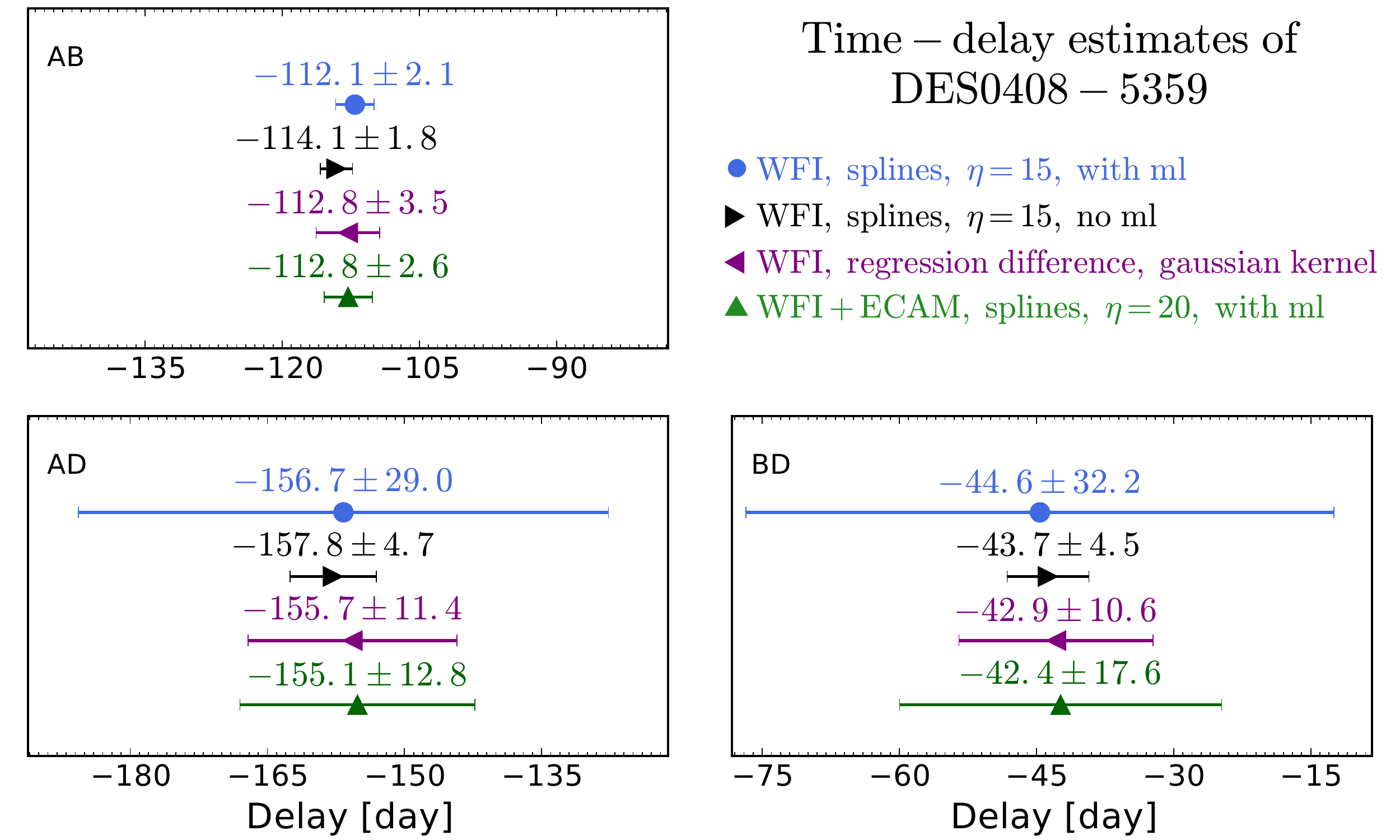}
\caption{{Time delay measurements for the 3 brightest quasar images of \obj\ using the data shown in Fig.~\ref{fig:lc}. The time delay measurements are carried out in 3 different ways. In blue are shown the results using only the WFI data and the spline fitting method and in purple with the regdiff method. In green are shown the results for the combined WFI+Euler data set using the spline fitting method. For comparison we also show, in black, the spline fitting result when using only the high-cadence WFI data and no model for the microlensing extrinsic variations. A negative $\Delta t(AB)$ means that signal from image A reaches the observer's plane before B.}}
\label{fig:delays}
\end{figure*}

We carry out the photometric measurements using the deconvolution photometry with the so-called MCS image deconvolution algorithm \citep{MCS,Cantale2016}. This algorithm first computes a deconvolution kernel from the images of stars. The kernel is chosen so that the Point Spread Function (PSF) in the deconvolved images is a circular Gaussian function with a Full-Width-Half-Maximum (FWHM) of 2 pixels. The pixel size in the deconvolved images is half that of the original data, i.e. the resolution in the resulting images is 0.2\arcsec\ for WFI and the pixel size is 0.1\arcsec. We show in Fig.~\ref{fig:field} the PSF stars used as well as the reference stars for the image-to-image flux calibration.

The MCS algorithm deconvolves all the registered images simultaneously, 
i.e. each one with its own PSF. However all images share the same 
deconvolved model, which is decomposed into a point-source channel 
(quasar images) and an extended-channel (lensing galaxy and faint quasar 
host galaxy). In this process, the position of the point sources is the 
same for all images as well as the extended-channel but the intensities 
of the point sources vary from image to image, hence leading to the 
photometric light curves. The latter are presented for the three 
brightest lensed images of \obj\ in Fig.~\ref{fig:lc} which shows the 
striking difference in depth and sampling between the 2.2m data and the 
Euler ones. Yet, the two data sets agree and complement each other well 
and exhibit fine structures in the light curves of A and B. The D 
component, however, has much shallower variations given the larger 
photometric error bars and will need further monitoring.

\section{Time delay measurement}
\label{sec:PyCS}

We use {\tt PyCS}\footnote{PyCS can be obtained from \url{http://www.cosmograil.org}}, a toolbox containing several algorithms to measure time delays from quasar light curves and accounting for the intrinsic variations (from the lensed source) and extrinsic variations (microlensing) in the data. The (public) algorithm was proposed by \citet{Tewes2013b} and tested on simulated data from the Time Delay Challenge \citep{BonvinPyCS2016, Liao2015}, with overall excellent performances.

\subsection{Time delay measurement with {\tt PyCS}}

{\tt PyCS} is the standard curve-shifting toolbox of the COSMOGRAIL 
project. We apply it to the WFI light curves as well as on the 
combination of the Euler and WFI data. We do not attempt to measure a 
time delay using the Euler data on their own as they contain only 45 
epochs over the duration of the observations. However, used in 
combination with the WFI images, they increase the time baseline with 
observations between July 2016 and October 2016. 

We use the two best algorithms of the {\tt PyCS} toolbox: the {\it 
free-knot spline} technique and the {\it regression difference} 
technique \citep{Tewes2013b}. In the former, the intrinsic and extrinsic 
variations in the light curves are modeled explicitly as spline 
functions. In doing so, we give more flexibility to the spline 
representing the intrinsic variations of the quasar than to the spline 
representing extrinsic variations. This flexibility is varied through a 
parameter $\eta$ representing the initial knot spacing in the splines 
\cite[][]{Tewes2013b}. The value of $\eta$ is optimized using simulated 
light curves that mimic the properties of the data. These simulated 
light curves are produced using the same toolbox as in 
Sect.~\ref{error}. Note, however, that our results depend little on the 
exact value for $\eta$ and that for a fixed choice of $\eta$ the 
position of the knots can change during the fit. This avoids many of the 
traditional "oscillation" problems with spline fit using fixed (and 
possibly badly placed) knots. To fit the present data, we use splines 
with only 3 knots to represent the microlensing. We further impose that 
the central knot stays centered on the temporal axis between the light 
curve extrema during the minimization process. Adding knots to the 
microlensing splines does not change significantly the results.

In the {\it regression difference} technique, we minimize the variability of the difference between Gaussian-process regressions performed on each light curve individually. This second method has no explicitly parametrized form for the extrinsic variability, which makes the two techniques fundamentally different and independent. We apply the two methods to our data in the same way as in \citet{Bonvin2017} and \citet{Tewes2013b}, who also give the procedure to derive the random and systematic errors from simulated light curves. As the results may depend on some of the key parameters that characterize each method, we perform robustness checks identical to the ones in
\citet{Bonvin2017}; we use {\tt PyCS} with a large range of method parameters, namely the number of  knots of the spline technique and the covariance function of the {\it regression difference} technique \citep[see][for a full description of these parameters]{Tewes2013b}. We do not find significant differences in the mean time-delay values among the results obtained for the various tests, although the precision might vary. 

\begin{figure*}[t!]
\centering
\includegraphics[width=0.98\textwidth]{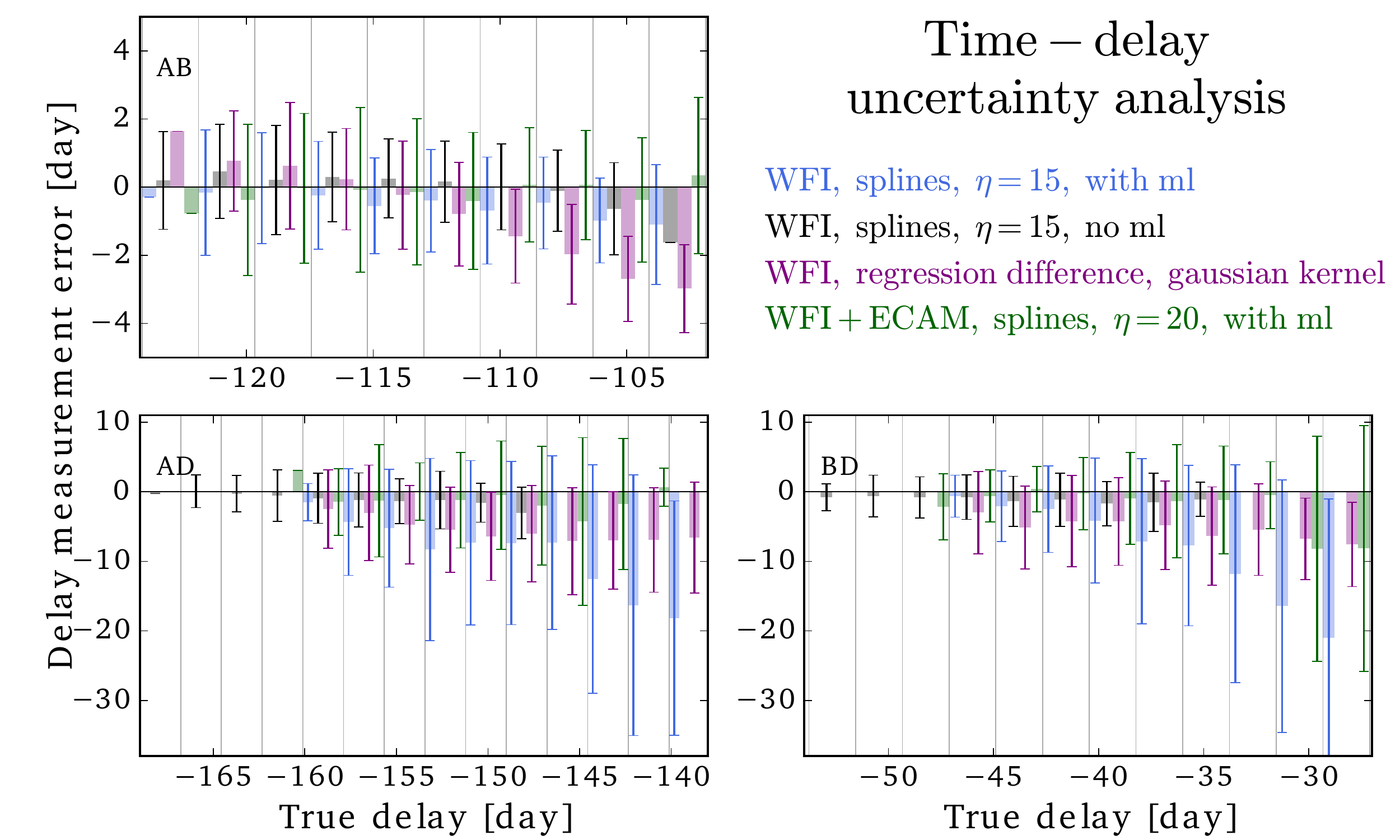}
\caption{{Error estimates for the time-delay measurements performed on 1000 simulated light curves. The color code is the same as in Fig.~\ref{fig:delays}. The $x$-axis of each panel shows the values for the true time-delay in the mock light curves. For each time-delay value are shown the random error bars as thin lines and the systematic errors as thick lines. The values for the time delays, as measured on the real data correspond to the center of each panel.}}
\label{fig:errors}
\end{figure*}

The time-delay measurements are summarized in Fig.~\ref{fig:delays} for different data sets, i.e. with and without the Euler data. We note that the longer-baseline Euler data improve the time-delay estimates involving the D image when used in combination with the 2.2m WFI data. The latter have strong constraining power as high-frequency structures are captured in the curves, i.e. mostly the A and B components which display two strong features, including two inflection points in the case of B. For the much fainter D image, the situation is more complex as it displays only one shallow inflection point and no clear feature that can be matched to the other light curves. 

\subsection{Error estimates}
\label{error}

In {\tt PyCS}, the error estimates are done by running the 
curve-shifting techniques on mock light curves created from a generative 
model \citep[see][]{Tewes2013b}. In these mocks, the intrinsic and 
extrinsic variations of the quasar are the same as the one inferred from 
the real data as well as the temporal sampling and photometric errors. 
What changes from mock to mock are the correlated extrinsic variability 
(whose statistical properties mimic the observations), the photometric 
noise, the true time delays and the value of the simulated data points. 
The mock curves are drawn so that they have the same ``time-delay 
constraining power" than the original data, i.e. the properties of the 
residuals after fitting the mock with a spline are statistically the 
same as on the real data. 

We carry out simulations for a broad range of true input time delays 
around the measured value. This error analysis is summarized in 
Fig.~\ref{fig:errors} which, for each of the true time delay tested, 
provides the random and systematic errors. The final error for the delay 
is taken as the worst random error over all the bins, combined in 
quadrature with the worst systematic error. Obviously, the size of the 
bins, as well as the range of true time-delays explored when drawing the 
simulated light curves can have an impact on the final error. Part of 
the robustness checks we performed are intended to ensure that we do not 
overestimate or underestimate the errors by choosing inappropriate bin 
sizes and ranges in true time-delays. In the present case, our choice of 
possible true time-delays ranges up to $\pm$10 days from our initial 
estimation obtained by running our point estimator on the original data. 
Such a wide range encompasses our uncertainty regarding the time delays 
of \obj\ that have never been measured before, yet is also small enough 
to make sure that simulated light curves (especially A and D) are 
sufficiently overlapping. We can note from the bottom panels of 
Fig.~\ref{fig:errors}, that the large systematic errors of the spline 
optimizer (in blue and green) come mostly from the simulations with 
extreme values of true delays. 



We note that for the A and B light curves, which benefit from high-S/N 
data, the time delay is very accurate, whatever true delay is tested.  
This illustrates the importance of catching as many faint and 
short-duration structures in the light curves and the impact of 
high-cadence and high-S/N data. As a robustness test, we carry out the 
time-delay measurements without modeling explicitly extrinsic variations 
when using the spline technique. These results are indicated in black in 
Figs.~\ref{fig:delays} \& \ref{fig:errors} and show that the value of 
the time delays do not depend much on the extrinsic variations for the 
AB delay. Future observations of other objects will show if this is 
specific to \obj\ or a more general behaviour of the results with 
high-cadence and high-S/N light curves.  Our final time delay value for 
AB is $\Delta t(AB)=-112.1 \pm 2.1$ days (1.8\%), as obtained with the 
{\it free-knot spline} method using only the WFI data and extrinsic 
variations explicitly included. We make this choice because the 
time-delay measurement is precisely determined mostly thanks to the 
finely modeled peak in the WFI light curves (around mhjd = 57710 in A 
and 57820 in B). This peak is only crudely visible in the Euler data. 
Thus, adding the latter data set in the present case would only increase 
the overall noise. We also note that including extrinsic variations 
explicitly only slightly shifts the result while keeping the precision 
unchanged, as shown in Fig.~\ref{fig:delays}. We chose nevertheless to 
explicitly include extrinsic variations since the residuals in the data 
speak in favor of it and since it is physically motivated: 
microlensing and the subsequent extrinsic variations is present at some 
level in almost every lensed quasar know to date \citep{Mosquera2011}. 

In contrast to AB, the precision on the AD and BD delays depends on how 
extrinsic variations are modeled. Due to the lack of fine and sharp 
structures in the D light curve, and with only one shallow inflection 
point, the intrinsic and extrinsic variations are almost fully 
degenerate. We choose as our final results for these two delays the 
values obtained with the {\it free-knot spline} techniques for the 
combined Euler and WFI data sets. These times delay estimates are 
$\Delta t(AD) = -155.5 \pm 12.8$ days (8.2\%) and $\Delta t(BD) = -42.4 
\pm 17.6$ days (41\%).

Finally, it is worth noting that the delays obtained with the {\it regression difference} technique applied only to WFI are consistent with the ones from the {\it free-knot spline} techniques, and the AD and BD delays are even more precisely measured. We however prefer to stick to the results of the {\it free-knot splines} since they are much more precise for AB, which is currently - and by far - the most constraining delay to be used in future modeling of this lensed system. We also explicitly avoid cherry-picking the best technique per delay, i.e. in the present case the {\it free-knot splines} for AB and the {\it regression difference} for AD and BD. Such an ad'hoc choice may introduce a bias difficult to quantify. Finally, the {\it free-knot splines} technique is the one giving the smallest systematics according to Fig.~\ref{fig:errors}.

\subsection{Comparison with simple models}
\label{models}

Simple lens models are provided by \citet{Agnello2017} and are useful to 
design the monitoring strategy. They show that \obj\ is challenging in 
terms of measuring time delays, as the longest one is half the 
visibility window of the object. Even under these difficult conditions, 
the MPIA 2.2m data allow us to measure it to 1.8\% accuracy and 
precision, by unveiling short and small photometric variations of the 
quasar images A and B. 




The time delay we find for AB is in marginal agreement with the predictions of \citet{Agnello2017} but a detailed comparison between our measurement and the model predictions for \obj\ would be hazardous at this stage. Indeed, two values of the delay are possible as quasar image C has a possible companion in its vicinity. The predicted delays are $\Delta t(AB) \sim -85$ days for a model with a companion and $\Delta t(AB) \sim -125$ days for a model without a companion galaxy, which introduces degeneracies in the models that cannot be lifted with the current imaging data, which only provide the relative positions of the quasar images relative to the lensing galaxy. Until deep HST or ground-based AO images are available it will impossible to discriminate between models with or without this companion object. 

Independently of the characterization of the companion, the minimum 
information required to constrain lens models, even with few degree of 
freedom, are 1- sharp images of the lensed quasar host galaxy and 2- 
dynamics of the main lensing galaxy, as is used for all objects in the 
H0LiCOW program \citep{Suyu2017, Bonvin2017}. \citet{Suyu2014} 
illustrate how a radially thick Einstein ring can lift most model 
degeneracies, in combination with the central velocity dispersion of the 
lens. Acquisition of both HST images of the Einstein ring and spectra of 
the lens are under way.

\section{Conclusion}

We demonstrate a new observational strategy for measuring time delays in lensed quasars, using high-cadence and high signal-to-noise monitoring photometry.  The data, obtained almost daily over 7 months with the MPIA 2.2m telescope at ESO La Silla allow us to measure $\Delta t(AB) = -112.1 \pm 2.1$ days (1.8\%), $\Delta t(AD) = -155.5 \pm 12.8$ days (8.2\%) and $\Delta t(BD) = -42.4 \pm 17.6$ days (41\%), where the error bars include systematics due to residual extrinsic variations. For the AB time delay, we note that the time-delay values depend little on the way extrinsic variations are modeled, hence indicating that the high-frequency signal in the light curves from the WFI instrument on the 2.2m telescope is dominated by the intrinsic variations of the quasar, as expected. For the D component, however, we only report a tentative delay due to the lack of fast variations seen in the light curve of this faint lensed image.

With the current imaging data for \obj\ it is too early to compare in 
detail the model predictions for the time delays \citep{Agnello2017} and 
our measurements. Imaging with the HST or with ground-based adaptive 
optics (AO) is mandatory before drawing any conclusions and before doing 
any cosmological inference with \obj. Fortunately, the object has 
several bright stars in its immediate vicinity, making it an excellent 
prey for VLT and AO, e.g. with the MUSE integral field spectrograph, 
providing the required dynamical information on the lensing galaxy(ies). 
In the near-IR observations with the VLT and the Hawk-I imager and the 
GRAAL AO system would both allow to measure the dynamics of the lens and 
probe the mass along the line of sight on a 7\arcmin $\times$ 7\arcmin\ 
field of view. 

With the many wide-field surveys taking place at the moment (DES, KiDS, 
CFIS, DECals, HSC) and with LSST and Euclid coming in a fairly near 
future, the available number of lensed quasars will increase 
dramatically \citep[e.g.][]{Oguri2010}. We show how daily and very high 
signal-to-noise observations during 1 single season can match and 
potentially surpass long-term monitoring carried out at a lower rate 
(e.g. weekly) over many years. This should be accounted for when 
planning synoptic surveys such as the LSST. Although the latter will 
definitely yield very high signal-to-noise images, the full benefit of 
monitoring data with a 8m telescope will be much enhanced in combination 
with a daily cadence or very close to a daily cadence. 

We are currently monitoring 4 objects since October 2016 with the MPIA 2.2m. From a preliminary analysis of these 4 targets, we anticipate that reliable (i.e. to a few percents) time delays will be measured for 3 of them. We show here our results for \obj, for which the observing season is finished. 

\begin{acknowledgements}
The authors would like to thank R. Gredel for his help in setting up the program at the MPIA 2.2m telescope. This work is supported by the Swiss National Science Foundation (SNSF). S.H. Suyu and D.C.Y. Chao thank the Max Planck Society for support through the Max Planck Research Group for SHS. T. Treu acknowledges support by the National Science Foundation through grant 1450141, by the Packard Foundation through a Packard Research Fellowship and by the UCLA Dean of Physical Sciences. K.Rojas is supported by Becas de Doctorado Nacional CONICYT 2017. T. Anguita and M. Chijani acknowledge support by proyecto FONDECYT 11130630 and by the Ministry for the Economy, Development, and Tourism’s Programa Inicativa Cient\'{i}fica Milenio through grant IC 12009, awarded to The Millennium Institute of Astrophysics (MAS). M. Tewes acknowledges support from the DFG grant Hi 1495/2-1. J. Garcia-Bellido is supported by the Research Project FPA2015-68048 [MINECO-FEDER], and the Centro de Excelencia Severo Ochoa Program SEV-2012-0249. C.D. Fassnacht acknowledges support from the National Science Foundation grant AST-1312329 and from the UC Davis Physics Department and Dean of Math and Physical Sciences.

Funding for the DES Projects has been provided by the U.S. Department of Energy, the U.S. National Science Foundation, the Ministry of Science and Education of Spain, the Science and Technology Facilities Council of the United Kingdom, the Higher Education Funding Council for England, the National Center for Supercomputing Applications at the University of Illinois at Urbana-Champaign, the Kavli Institute of Cosmological Physics at the University of Chicago, the Center for Cosmology and Astro-Particle Physics at the Ohio State University, the Mitchell Institute for Fundamental Physics and Astronomy at Texas A\&M University, Financiadora de Estudos e Projetos, Funda{\c c}{\~a}o Carlos Chagas Filho de Amparo {\`a} Pesquisa do Estado do Rio de Janeiro, Conselho Nacional de Desenvolvimento Cient{\'i}fico e Tecnol{\'o}gico and the Minist{\'e}rio da Ci{\^e}ncia, Tecnologia e Inova{\c c}{\~a}o, the Deutsche Forschungsgemeinschaft and the Collaborating Institutions in the Dark Energy Survey. 

The Collaborating Institutions are Argonne National Laboratory, the University of California at Santa Cruz, the University of Cambridge, Centro de Investigaciones Energ{\'e}ticas, Medioambientales y Tecnol{\'o}gicas-Madrid, the University of Chicago, University College London, the DES-Brazil Consortium, the University of Edinburgh, the Eidgen{\"o}ssische Technische Hochschule (ETH) Z{\"u}rich, Fermi National Accelerator Laboratory, the University of Illinois at Urbana-Champaign, the Institut de Ci{\`e}ncies de l'Espai (IEEC/CSIC), the Institut de F{\'i}sica d'Altes Energies, Lawrence Berkeley National Laboratory, the Ludwig-Maximilians Universit{\"a}t M{\"u}nchen and the associated Excellence Cluster Universe, the University of Michigan, the National Optical Astronomy Observatory, the University of Nottingham, The Ohio State University, the University of Pennsylvania, the University of Portsmouth, SLAC National Accelerator Laboratory, Stanford University, the University of Sussex, Texas A\&M University, and the OzDES Membership Consortium.

The DES data management system is supported by the National Science Foundation under Grant Number AST-1138766. The DES participants from Spanish institutions are partially supported by MINECO under grants AYA2015-71825, ESP2015-88861, FPA2015-68048, SEV-2012-0234, SEV-2012-0249, and MDM-2015-0509, some of which include ERDF funds from the European Union. IFAE is partially funded by the CERCA program of the Generalitat de Catalunya.

\end{acknowledgements}

\bibliographystyle{aa}
\bibliography{paper}

\clearpage

\end{document}